\documentclass[preprint,preprintnumbers,aps,superscriptaddress,nofootinbib,tightenlines,floatfix]{revtex4}

\usepackage{graphicx}
\usepackage{bm}
\usepackage{comment}
\usepackage{color}

\newcommand{\gsim}{\raisebox{-0.7ex}{$\stackrel{\textstyle >}{\sim}$ }}
\newcommand{\lsim}{\raisebox{-0.7ex}{$\stackrel{\textstyle <}{\sim}$ }}

\def\OMIT#1{{}}

\def\siii{^3 \hskip -0.025in S _1}

\newcount\hour \newcount\hourminute \newcount\minute 
\hour=\time \divide \hour by 60
\hourminute=\hour \multiply \hourminute by 60
\minute=\time \advance \minute by -\hourminute
\newcommand{\mydate}{\ \today \ - \number\hour :\number\minute}

\begin{document}

\preprint{NT@UW-11-19}

\begin{figure}[!t]

  \vskip -1.5cm
  \leftline{\includegraphics[width=0.25\textwidth]{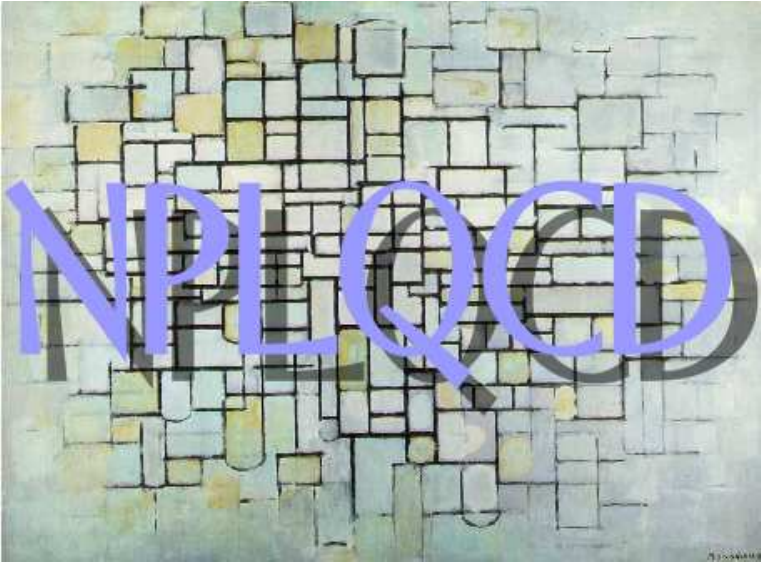}}
\end{figure}

\title{Improving the Volume Dependence of Two-Body Binding Energies Calculated with
  Lattice QCD}

\author{Zohreh Davoudi and Martin J.~Savage}
\affiliation{Department of Physics, University of Washington, Seattle,
  WA 98195-1560.}  

\date{\mydate}

\begin{abstract}
\noindent
Volume modifications to the binding of  two-body systems
in large cubic volumes of extent $L$
depend  upon the total  momentum
and exponentially upon the ratio of $L$ to the 
size of the boosted system.
Recent work by Bour {\it et al} determined the momentum dependence 
of the leading volume modifications
to nonrelativistic systems
with periodic boundary conditions imposed on the single-particle
wavefunctions, 
enabling them to numerically determine the scattering of such
bound states using a low-energy effective field theory and L\"uscher's
finite-volume method.
The calculation of bound nuclear systems directly from QCD using Lattice QCD
has begun, and it is important to reduce the systematic uncertainty
introduced into such calculations by the finite spatial extent of the
gauge-field configurations.
We extend the work of Bour {\it et al} from nonrelativistic quantum mechanics
to quantum field theory by generalizing the work of L\"uscher  and of Gottlieb
and Rummukainen to boosted two-body bound states.
The volume modifications to binding energies
can be exponentially reduced 
from ${\cal O}\left(e^{-\kappa L}/L\right)$ to ${\cal O}\left(e^{-2 \kappa L}/L\right)$
in nonrelativistic systems (where $\kappa$ is the binding momentum of the state)
by forming particular combinations of
the binding energies determined in the four lowest-lying boosted systems. 
Relativistic  corrections to this combination, and others,  that violate the
exponential reduction are determined.
An analysis of what can be expected from Lattice QCD calculations of
the deuteron is performed, the results of which are representative of a generic loosely bound system.
\end{abstract}
\pacs{}
\maketitle
%

\section{Introduction}
\noindent 
One of the major goals of nuclear physics research is to determine the properties and
interactions of nucleons and nuclei, and more generally hadrons, 
directly from the underlying theory of the strong interactions, 
quantum chromodynamics (QCD).
The only known way to accomplish this is to numerically evaluate the QCD
path-integral using  Lattice QCD (LQCD) in which the space-time continuum is replaced
by a finite-volume grid, and the integrals over the fields at each point in
space-time are performed using Monte-Carlo.  
The systematic errors introduced by a finite space-time volume, $L^3\times T$,
with a  finite
lattice spacing, $b$, can be systematically removed by performing calculations in multiple
volumes, with multiple lattice spacings and using the theoretical knowledge of
the associated functional dependences.   
The computational resources that are required for LQCD calculations 
with light quark masses ($m_q$) near their physical values, 
with small enough lattice spacings ($b\Lambda_\chi\ll 1$ where $\Lambda_\chi$
is the scale of chiral symmetry breaking),
and in large enough lattice volumes ($m_\pi L \gsim 2\pi$)
that permit reliable extrapolations and interpolations 
of phenomenologically important observables are now becoming available with the
deployment of multi-peta-flop machines.

While calculating (post-dicting) the masses of the lowest-lying hadrons 
to high precision is computationally demanding (even without
strong isospin breaking and electromagnetism),
calculating the binding energies of systems composed of the lowest-lying
hadrons (nuclei) to the same level of precision is significantly more demanding.
For instance, calculating the mass of the proton at the percent level requires
precision $\sim \pm 10~{\rm MeV}$, while calculating the deuteron
binding energy at the percent level requires
precision of $\sim \pm 20~{\rm keV}$.  So while the same gauge-field
configurations can be used for both calculations, the statistical precision
required in
the evaluation of the relevant correlation functions differs by orders of magnitude.
Further, while the error in the calculation of a hadron mass introduced by
the finite lattice volume scales as $\sim e^{-m_\pi L}$ for large volumes~\cite{Luscher:1985dn}, 
the error in the calculation of a binding energy is largely dictated by the
size of the bound state~\cite{Luscher:1985dn,Beane:2003da,Sasaki:2006jn}.  
For instance, for a two-body bound state, the finite
volume error scales as $\sim e^{-\kappa_0 L}/L$, where $\kappa_0$ is the binding
momentum of the state, $B=-\left(\sqrt{m_1^2-\kappa_0^2} +
  \sqrt{m_2^2-\kappa_0^2} - m_1-m_2\right)$, 
where $B$ is the infinite-volume binding
energy.   
For a loosely bound system such as the deuteron, 
the deviation between the ground-state energy calculated in a 
given LQCD calculation and the actual binding energy will
be dominated by the size of the deuteron~\footnote{The modifications due to the non-zero range
  of the nuclear forces scale as $\sim e^{-m_\pi L}$.} and will, in general, be significant
except in very large volumes.
These volume modifications arise from the exclusion of single-particle momentum
modes in the bound state wavefunction due to the periodic boundary conditions
(BC)  that are imposed on the quark and gluon fields in the spatial directions.

During the last year, LQCD calculations have observed bound systems of baryons.
Evidence for a bound H-dibaryon (a state with the quantum numbers of
$\Lambda\Lambda$) was found
in $n_f=2+1$ LQCD calculations with $m_\pi\sim 390~{\rm MeV}$~\cite{Beane:2010hg,Beane:2011xf}, and subsequent
evidence was found in $n_f=3$ calculations with $m_\pi\sim 840~{\rm
  MeV}$~\cite{Inoue:2010es}.
Also, evidence was reported for $^3$He, $^4$He~\cite{Yamazaki:2009ua} and the
deuteron~\cite{Yamazaki:2011nd} in quenched calculations with $m_\pi\sim 800~{\rm MeV}$.
The infinite volume binding energy for each of these
nuclei was determined  by calculating the ground state energy in a number ensembles
of gauge-fields with different volumes and then extrapolating to infinite volume.  
While the generation of quenched gauge-fields is inexpensive
computationally  compared to the generation of QCD gauge-fields,
the results of quenched calculations cannot be 
used to reliably predict quantities in nature.

An important observation that was recently made by Bour {\it et
  al}~\cite{Bour:2011ef} 
in nonrelativistic systems
is that
the volume modifications depend upon the momentum of the bound state in the
lattice volume, as moving bound states have different momentum modes
excluded from their two-body wavefunction.
The implication of this, when extended to quantum field theory, is that the 
unextrapolated
binding energies of composite systems, and in particular light nuclei,
calculated with LQCD will depend upon the total momentum of the system.
Bour {\it et al}~\cite{Bour:2011ef} were interested in calculating the
scattering of bound states
in a numerical evaluation of a low-energy effective field theory path integral, 
and found that the momentum dependent contribution to the two-body ground state
energy had to be removed prior to  using L\"uscher's
method~\cite{Luscher:1986pf,Luscher:1990ux} to determine the phase-shift.

In this work, we extend the quantum field theory formalism established 
by L\"uscher~\cite{Luscher:1986pf,Luscher:1990ux} and generalized to moving
systems by Rummukainen and Gottlieb~\cite{Rummukainen:1995vs},
to determine the volume modifications of binding
energies of bound systems composed of two spinless particles with s-wave interactions
moving in a finite cubic volume, extending the results obtained by Bour {\it et
  al}~\cite{Bour:2011ef}
in nonrelativistic quantum mechanics.
By forming combinations of the ground state energies of two-body systems
with different lattice momenta, volume modifications
can be exponentially suppressed in the nonrelativistic limit, and 
we determine the violations of this exponential suppression.
In the case of the deuteron, the lowest energy-eigenvalue in
the np system with lattice momenta of ${\bf P}={2\pi\over L}{\bf d}$ with
$|{\bf d}|^2=0,1,2,3$ allow for the infinite-volume deuteron binding energy to be
determined with a volume modification of  $\Delta B^{(vol)}\lsim 20~{\rm keV}$
for $L\gsim 12~{\rm fm}$, orders of magnitude smaller than in the
ground state energy of the $|{\bf d}|=0$ system alone.
From a practical standpoint, 
forming such linear combinations of boosted ground state
energies requires significantly smaller  computational resources than computing the
ground state energies  in multiple lattice volumes.  
The former can be
accomplished with one set of quark propagators
on one ensemble of gauge fields,
while the later requires generating multiple ensembles of gauge fields and
quark propagators on each.
In some sense, these linear combinations of binding energies represent an  exponential
``volume-improvement'' of the binding energy calculation on a given ensemble of
gauge-fields.

An analysis of possible determinations of the deuteron binding energy from a
single lattice volume was carried out in Ref.~{\cite{Beane:2010em}}.
For a large enough lattice volume, the energies of the lowest two levels 
with $|{\bf d}|=0$
fall below the t-channel cut, and the effective range expansion (ERE) of $p\cot\delta$ 
can be used.  Truncating the ERE at the first two terms, which is known to be  a good
approximation for nucleon-nucleon scattering, allows the
deuteron binding energy to be obtained by an interpolation.
Of course, it is desirable to not make such an approximation in extracting the
deuteron binding energy, but this would require more states below the t-channel
cut and hence even larger lattice volumes.

\section{Spinless Particles with S-Wave Interactions }
\noindent 
For spinless particles of mass $m_1$ and $m_2$ interacting in an s-wave,
the scattering amplitude below the inelastic threshold is uniquely specified by
the phase-shift $\delta_0$, and is proportional to $1/(p\cot\delta_0- i p)$
where $p$ is the magnitude of the momentum of each particle in the
center-of-momentum (CoM) frame.  
When these two particles are confined to a cubic volume of spatial extent $L$
subject to periodic BC's, and with total momentum 
${\bf P} = {2\pi\over L}{\bf  d}$, 
the energy-eigenvalues of the system are the solutions to~\footnote{
Exponentially suppressed corrections to this relation 
for $\pi\pi$ and $NN$ scattering
of the form $e^{-m_\pi L}$
have been determined  in Ref.~\cite{Bedaque:2006yi} and
Ref.~\cite{Sato:2007ms}, respectively.}
\begin{eqnarray}
q^*\cot\delta(q^*) & = & 
{2\over \gamma L \sqrt{\pi}}\ Z_{00}^{({\bf d})}(1;\tilde q^{*2},
\tilde \Delta m_{12}^2)
\ \ \ ,
\label{eq:evaluerel}
\end{eqnarray}
where 
\begin{eqnarray}
Z_{LM}^{({\bf d})}
& = & 
\sum_{\bf r}
{ |{\bf r}|^L \ Y_{LM}(\Omega_{\bf r})
\over
|{\bf r}|^2 - \tilde q^{*2}}
\ \ ,\ \ 
{\bf r} \ =\ {1\over \gamma}\left( {\bf n}_\parallel +\alpha {\bf d} \right) +
{\bf n}_\perp
\ =\ \hat\gamma^{-1} \left( {\bf n} +\alpha {\bf d} \right)
\ \ \ ,
\label{eq:Zdef}
\end{eqnarray}
where  ${\bf n}$ is a triplet of integers.
A `` * '' denotes a quantity determined in the CoM frame, and for a system with
${\bf P} = {2\pi\over L}{\bf  d}$ and total energy $E$, the energy in the CoM
frame is $s = E^{*2}=E^2-|{\bf P}|^2$ which defines the $\gamma$-factor (which
depends explicitly on ${\bf d}$), 
$\gamma=E/E^*$.  The magnitude of the three-momentum in the CoM frame, $q^*$,
is determined by
$E^{*2}=(\sqrt{q^{*2}+m_1^2} + \sqrt{q^{*2}+m_2^2})^2$ and 
the factor $\alpha$ that appears in eq.~(\ref{eq:Zdef}) is 
\begin{eqnarray}
\alpha\ =\ {1\over 2}\left[1\ +\ {m_1^2-m_2^2\over E^{*2}}\right]
\ \ \ .
\label{eq:alphadef}
\end{eqnarray}
In eq.~(\ref{eq:evaluerel}), $\Delta m_{12}^2$ is defined as  
$\Delta m_{12}^2= m_1^2-m_2^2$, and a  tilde over any variable denotes scaling by a factor
of $L/(2\pi)$, e.g. $\tilde q^* = q^* L/(2\pi)$.
In the case  of equal masses ($m_1=m_2$) $\alpha={1\over 2}$, and 
the expressions in eq.~(\ref{eq:evaluerel}) and
eq.~(\ref{eq:Zdef}) 
reduce to  the known result for boosted systems of equal
mass~\cite{Rummukainen:1995vs, Kim:2005gf,Christ:2005gi}.
Further developments are required in order
to recover the results obtained for nonrelativistic systems by Bour {\it et al}
from eq.~(\ref{eq:evaluerel}), eq.~(\ref{eq:Zdef}) and eq.~(\ref{eq:alphadef}),
as we now describe.

Assuming the scattering amplitude admits a single bound state in infinite
volume,  the location of the lowest energy-eigenvalue in a finite cubic
volume is dictated by the behavior of  $Z_{LM}^{({\bf d})}$ for $\tilde q^{*2} <0$.
It is clear from the form of $Z_{LM}^{({\bf d})}$ in eq.~(\ref{eq:Zdef}) that
there are no poles along the negative axis, and the Poisson resummation
formula can be used to determine its asymptotic behavior at large $-\tilde
q^{*2}$.
It is straightforward to show that for $q^{*2} < 0$
\begin{eqnarray}
Z_{00}^{({\bf d})} (1;\tilde q^{*2},
\tilde \Delta m_{12}^2)
& \rightarrow  & 
{\gamma\over\sqrt{4\pi}} 
\left[\ 
-2\pi^2
\sqrt{-\tilde q^{*2}}
\ +\ 
\sum_{ {\bf m}\ne {\bf 0} }\
{\pi\over |\hat\gamma {\bf m}|}\ 
e^{i 2\pi \alpha {\bf m}\cdot {\bf d}}\ 
e^{-2\pi |\hat\gamma {\bf m}| \sqrt{-\tilde q^{*2}}  }
\right]
\ \ \ ,
\end{eqnarray}
where
\begin{eqnarray}
\hat\gamma {\bf m} & = & \gamma {\bf m}_\parallel + {\bf m}_\perp
\ =\ 
(\gamma-1) {{\bf m}\cdot {\bf d}\over |{\bf d}|^2} \ {\bf d} \ +\ 
{\bf m}
\ \ \ ,
\end{eqnarray}
and the ${\bf m}$ are  triplets of integers.
By setting $q^*=i\kappa$, 
the eigenvalue equation in eq.~(\ref{eq:evaluerel})
becomes
\begin{eqnarray}
p\cot\delta(p)\big|_{p=i\kappa}
\ +\ \kappa
 & = & 
{1\over L}\ 
\sum_{ {\bf m}\ne {\bf 0} }\
{1\over |\hat\gamma {\bf m}|}\ 
e^{i 2\pi \alpha {\bf m}\cdot {\bf d}}\ 
e^{- |\hat\gamma {\bf m}| \kappa L  }
\ =\ 
{1\over L} \ F^{({\bf d})}(\kappa L)
\ \ \ .
\label{eq:pcotkappa}
\end{eqnarray}
In the infinite volume limit $F^{({\bf d})}(\kappa L)=0$, and the eigenvalue
equation becomes
\begin{eqnarray}
p\cot\delta(p)\big|_{p=i\kappa_0}
\ +\ \kappa_0
 & = & 
0
\ \ \ ,
\end{eqnarray}
which correctly reproduces the location of the pole in the S-matrix.
While it is not required for this analysis, below the t-channel cut 
$p\cot\delta(p)$ can be expanded in powers of the CoM energy 
$p\cot\delta(p)=-{1\over a} + {1\over 2} r p^2+...$, defining the 
ERE, where $a$ is the scattering length and $r$ is the
effective range.
For the lowest few ${\bf d}$ vectors, the finite volume 
functions $F^{({\bf d})}(\kappa L)$ are (where we keep in mind that $\gamma$
depends upon $|{\bf d}|$)
\begin{eqnarray}
F^{(0,0,0)}(\kappa L) 
& = & 
6 \ e^{-\kappa L }
\ +\ 
6 \sqrt{2}\ e^{-\sqrt{2}\kappa L }
\ +\ 
{8\over\sqrt{3}}\ e^{-\sqrt{3}\kappa L }
\ +\ 
3 \ e^{-2 \kappa L }
\ +\ ...
\nonumber\\
F^{(0,0,1)}(\kappa L) 
& = & 
2 \left(\ 2 \ e^{-\kappa L}\ 
\ +\ 
{\cos\left(2\pi\alpha\right)\over\gamma} e^{-\gamma \kappa L }\ \right)
\nonumber\\
&& \ +\ 
2\sqrt{2} 
\left(\ e^{-\sqrt{2}\kappa L }
\ +\ 
2 \cos\left(2\pi\alpha\right) \sqrt{{2\over\gamma^2+1}}\ e^{-\sqrt{\gamma^2+1}\kappa L } 
\right)
\nonumber\\
&& 
\ + \ 
{8\cos\left(2\pi\alpha\right) \over\sqrt{\gamma^2+2}}  e^{-\sqrt{\gamma^2+2}\kappa L } 
\ +\ 
\left(\ 2 \ e^{-2\kappa L}\ 
\ +\ {\cos\left(4\pi\alpha\right) \over\gamma} e^{-2 \gamma \kappa L }\ \right)
\ +\ ...
\nonumber\\
F^{(0,1,1)}(\kappa L) 
& = & 
2\left(\ e^{-\kappa L}\ 
\ + \ 2 \cos\left(2\pi\alpha\right) \sqrt{2\over\gamma^2+1} e^{-\sqrt{\gamma^2+1\over 2} \kappa L}\ \right)
\nonumber\\
& & \ +\ 
\sqrt{2}\ \left(\ 
e^{-\sqrt{2}\kappa L }
\ +\ 
{\cos\left(4\pi\alpha\right)\over\gamma }\  e^{-\sqrt{2}\gamma \kappa L }
\ +\ 
{8 \cos\left(2\pi\alpha\right) \over\sqrt{3+\gamma^2}}\  e^{-\sqrt{3+\gamma^2\over 2}\kappa L }
\ \right)
\nonumber\\
& & \ +\ 
{4\over\sqrt{3}}\left(\ 
e^{-\sqrt{3}\kappa L }
\ +\ 
\cos\left(4\pi\alpha\right) \sqrt{3\over 2\gamma^2+1} e^{-\sqrt{2\gamma^2+1}\kappa L }
\right)
\nonumber\\
& & \ +\ 
\left(\ 
\ e^{-2 \kappa L}\ 
\ +\ 2 \cos\left(4\pi\alpha\right) \sqrt{2\over\gamma^2+1} e^{-2 \sqrt{\gamma^2+1\over 2} \kappa L}\ \right)
+\ ...
\nonumber\\
F^{(1,1,1)}(\kappa L) 
& = & 
6 \cos\left(2\pi\alpha\right)  \sqrt{3\over\gamma^2+2} e^{-\sqrt{\gamma^2+2\over 3} \kappa L}
\nonumber\\
& & \ +\ 
3\sqrt{2} 
\left(\ 
e^{-\sqrt{2}\kappa L }
\ +\  \cos\left(4\pi\alpha\right)
\sqrt{3\over 2\gamma^2+1}\  e^{-\sqrt{{2\over 3}(2\gamma^2+1)}\kappa L }
\ \right)
\nonumber\\
& & 
\ + \ 
{2\over\sqrt{3}} 
\left(
{\cos\left(2\pi\alpha\right)\over\gamma } e^{-\sqrt{3}\gamma \kappa L }
\ +\ \cos\left(6\pi\alpha\right)
{9\over\sqrt{\gamma^2+8}} e^{-\sqrt{\gamma^2+8\over 3}\kappa L }
\right)
\nonumber\\
& & \ +\ 3 \cos\left(4\pi\alpha\right)
\sqrt{3\over\gamma^2+2} e^{-2 \sqrt{\gamma^2+2\over 3} \kappa L}
\ +\ ...
\ \ \ ,
\label{eq:Fdef}
\end{eqnarray}
where the ellipses denotes terms that scale as $\sim e^{-\sqrt{5}\kappa_0 L}/L$
and higher.
In the large volume limit where the $F^{({\bf d})}$ will give rise to $\kappa^{({\bf d})}$
that are  close to $\kappa_0$, the 
$\kappa^{({\bf d})}$ can be determined in a perturbative solution to eq.~(\ref{eq:pcotkappa}).
Introducing the dimensionless parameter $\lambda$, writing $\kappa^{({\bf d})}=\kappa_0
+ \lambda\kappa_1^{({\bf d})}+\lambda^2\kappa_2^{({\bf d})}+...$ 
along with the substitution $F^{({\bf d})}(\kappa^{({\bf d})} L) \rightarrow
\lambda F^{({\bf d})} (\kappa^{({\bf d})} L)$ and equating orders in $\lambda$, leads to
\begin{eqnarray}
\kappa_1^{({\bf d})} & = & 
\ {Z_\psi^2\over L}\ F^{({\bf d})} (\kappa_0 L)
\nonumber\\
\kappa_2^{({\bf d})} & = & 
Z_\psi^2\ 
\left(\ 
\kappa_1 {1\over L} {d\over d\kappa} F^{({\bf d})}(\kappa_0 L)
\ +\ 
\kappa_1^2 {d\over dp^2}p\cot\delta\big|_{i\kappa_0}
\ -\ 
2\kappa_0^2\kappa_1^2
{d^2\over (dp^2)^2}p\cot\delta\big|_{i\kappa_0}
\right)
\nonumber\\
Z_\psi & = & {1\over\sqrt{1 - 2\kappa_0 {d\over dp^2}p\cot\delta\big|_{i\kappa_0}}}
\ , 
\label{eq:Zkappa}
\end{eqnarray}
where $Z_\psi^2$ is the residue of the bound-state pole in the S-matrix,
and the higher $\kappa_i^{({\bf d})}$ can be determined in a similar way.
For a given boost-vector, $\kappa_1^{({\bf d})}$ scales as $\kappa_1^{({\bf d})}\sim e^{-\kappa_0 L}/L$ and 
$\kappa_2^{({\bf d})}$ scales as $\kappa_2^{({\bf d})}\sim e^{-2 \kappa_0 L}/L$.  Consequently, the
contributions to $\kappa^{({\bf d})}$ that scale as $\sim e^{-\kappa_0 L}/L$,  $\sim
e^{-\sqrt{2}\kappa_0 L}/L$ and  $\sim e^{-\sqrt{3}\kappa_0 L}/L$  originate in
$\kappa_1^{({\bf d})}$ and are of the forms given in eq.~(\ref{eq:Fdef}).

It is clear from the explicit expressions for $F^{({\bf d})}(\kappa L)$ given
in eq.~(\ref{eq:Fdef}) that linear combinations that provide universal
cancellations of finite volume effects in binding energies do not exist in general. 
This is due to the appearances of both $\gamma$-factors 
and $\alpha$-factors that explicitly depend upon ${\bf d}$.
However, the nonrelativistic (NR) limit where $\gamma=\gamma^{({\rm NR})}=1$ and neglecting the
binding energy compared to the rest mass of the constituent hadrons,
$\alpha=\alpha^{({\rm NR})}=m_1/(m_1+m_2)$,
allows  for relations to be constructed for a given value of 
$\alpha^{({\rm NR})}$. 
Corrections to the relations can then be constructed as an expansion in 
$\gamma-\gamma^{({\rm NR})}$ and $\alpha-\alpha^{({\rm NR})}$.
The case of equal masses, $m_1=m_2$, is special because $\alpha={1\over 2}$
for any binding energy and not just in the limit where the binding energy is
small compared to the rest masses.
In the NR  limit it is useful to write 
\begin{eqnarray}
F^{({\bf d})}(\kappa L) & \rightarrow & \sum_j\ f_j^{({\bf d})( \alpha^{({\rm NR})})}\
{e^{-\sqrt{j}\kappa L}\over\sqrt{j}}
\ , 
\label{eq:Fexp}
\end{eqnarray}
where the coefficients $f_j^{({\bf d})( \alpha^{({\rm NR})})}$ for the case of equal
masses, $m_1=m_2$ with $\alpha^{({\rm NR})}={1\over 2}$, are given in 
Table~\ref{table:alpha2}
and the case where $m_2=2 m_1$ with $\alpha^{({\rm NR})}={1\over 3}$ is given
in 
Table~\ref{table:alpha3}.
\begin{table}[!ht]
  \caption{The coefficients $f_j^{({\bf d})({1\over 2})}$ in eq.~(\ref{eq:Fexp}) 
that determine the leading four finite-volume corrections to the two-body binding energy
in the nonrelativistic limit for a system with $m_1=m_2$.
  }
  \label{table:alpha2}
  \begin{ruledtabular}
    \begin{tabular}{c||cccc}
      $|{\bf d}|^2$ \qquad &  $f_1^{({\bf d})({1\over 2})}$  &   $f_2^{({\bf
          d})({1\over 2})}$ &  
$f_3^{({\bf d})({1\over 2})}$ &
       $f_4^{({\bf d})({1\over 2})}$ \\
      \hline
      $0$ & $6$  &  $12$  &  $8$ &  $6$ \\
      $1$ & $2$  &  $-4$  &  $-8$ &  $6$ \\
      $2$ & $-2$  &  $-4$  &  $8$ &  $6$ \\
      $3$ & $-6$  &  $12$  &  $-8$ &  $6$ \\
      \hline
    \end{tabular}
  \end{ruledtabular}
\end{table}
\begin{table}[!ht]
  \caption{The coefficients $f_j^{({\bf d})({1\over 3})}$ 
in eq.~(\ref{eq:Fexp}) 
that determine the leading four finite-volume corrections to the two-body binding energy
in the nonrelativistic limit for a system with $m_2=2m_1$.
  }
  \label{table:alpha3}
  \begin{ruledtabular}
    \begin{tabular}{c||cccc}
      $|{\bf d}|^2$ \qquad &  $f_1^{({\bf d})({1\over 3})}$  &   $f_2^{({\bf
          d})({1\over 3})}$ &  
$f_3^{({\bf d})({1\over 3})}$ &
       $f_4^{({\bf d})({1\over 3})}$ \\
      \hline
      $0$ & $6$  &  $12$  &  $8$ &  $6$ \\
      $1$ & $3$  &  $0$  &  $-4$ &  $3$ \\
      $2$ & $0$  &  $-3$  &  $2$ &  $0$ \\
      $3$ & $-3$  &  $3$  &  $5$ &  $-3$ \\
      \hline
    \end{tabular}
  \end{ruledtabular}
\end{table}
The ratios of coefficients in $f_1^{({\bf d})({1\over 3})}$ column in
Table~\ref{table:alpha3} reproduce the quantity $\tau({\bf k},{1\over 3})$
that are tabulated in Table 1 in Ref.~\cite{Bour:2011ef}.

It is also worth pointing out that the ground state energy of the unboosted, $|{\bf d}|=0$,
equal-mass, $\alpha={1\over 2}$, 
system
has leading and subleading volume corrections that are three times larger
than those of the $|{\bf d}|=1$ and $|{\bf d}|=\sqrt{2}$ systems.  So while it does
not constitute an exponential reduction in the volume modifications, the
binding energy
of the $|{\bf d}|=1$ and $|{\bf d}|=\sqrt{2}$ systems will be significantly closer to
the infinite volume binding energy than that of the  $|{\bf d}|=0$ system.

\subsection{Volume-Improvement for Equal Mass Systems :   $\alpha={1\over 2}$}
\noindent
The equal mass systems are special, as mentioned previously, because
$\alpha={1\over 2}$ is independent of the binding energy of the system, a
feature that is not present for $m_1\ne m_2$.
Using the coefficients in Table~\ref{table:alpha2}
it is straightforward to construct relations that eliminate the leading and
subleading orders of the volume modifications.  
We consider five relations:
\begin{eqnarray}
\overline{\kappa}^A & = & 
{1\over 8}\left(\ 
\kappa^{(0,0,0)} \ +\  
3 \kappa^{(0,0,1)} \ +\  
3\kappa^{(0,1,1)} \ +\  
\kappa^{(1,1,1)}\ 
\right)
\nonumber\\
& = & \kappa_0\ +\ 
{3 Z_\psi^2\over 2 L} \ \eta^2 \ \left(1+\kappa_0 L\right) e^{-\kappa_0 L}
\ +\ 
{\cal O}\left( \eta^4 e^{-\kappa_0 L} L, {e^{-2\kappa_0 L}\over 2 L}\right) 
\nonumber\\
\overline{\kappa}^B & = & 
{1\over 4}\left(\ 
\kappa^{(0,0,0)} \ +\  
3\kappa^{(0,1,1)} 
\right)
\nonumber\\
& = & \kappa_0\ +\ 
{3 Z_\psi^2\over 2 L} \eta^2 \ \left(1+\kappa_0 L\right) e^{-\kappa_0 L}
\ +\ 
{\cal O}\left( \eta^4 e^{-\kappa_0 L} L, {e^{-\sqrt{3} \kappa_0 L}\over \sqrt{3} L}\right) \nonumber\\
\overline{\kappa}^C & = & 
{1\over 4}\left(\ 
\kappa^{(1,1,1)} \ +\  
3\kappa^{(0,0,1)} 
\right)
\nonumber\\
& = & \kappa_0\ +\ 
{3 Z_\psi^2\over 2 L} \eta^2 \ \left(1+\kappa_0 L\right) e^{-\kappa_0 L}
\ +\ 
{\cal O}\left( \eta^4 e^{-\kappa_0 L} L, {e^{-\sqrt{3} \kappa_0 L}\over \sqrt{3} L}\right) 
\nonumber\\
\overline{\kappa}^D & = & 
{1\over 4}\left(\ 
\kappa^{(0,0,0)} \ +\  
\kappa^{(0,0,1)} \ +\  
\kappa^{(0,1,1)} \ +\  
\kappa^{(1,1,1)}\ 
\right)
\nonumber\\
& = & \kappa_0\ +\ 
{3 Z_\psi^2\over 2 L} \ \eta^2 \ \left(1+\kappa_0 L\right) e^{-\kappa_0 L}
\ +\ 
{\cal O}\left( \eta^4 e^{-\kappa_0 L} L, {e^{-\sqrt{2}\kappa_0 L}\over \sqrt{2} L}\right) 
\nonumber\\
\overline{\kappa}^E & = & 
{1\over 2}\left(\ 
3\kappa^{(0,0,1)} \ -\   
\kappa^{(0,0,0)}\ 
\right)
\nonumber\\
& = & \kappa_0\ +\ 
{3 Z_\psi^2\over 2 L} \ \eta^2 \ \left(1+\kappa_0 L\right) e^{-\kappa_0 L}
\ +\ 
{\cal O}\left( \eta^4 e^{-\kappa_0 L} L, {e^{-\sqrt{2}\kappa_0 L}\over \sqrt{2} L}\right) 
\ , 
\label{eq:twobodyana}
\end{eqnarray}
where $\gamma^2 = 1 + \eta^2 |{\bf d}|^2$ with
$\eta = {2\pi\over L E^*}$.
From the forms of the volume expansions given in eq.~(\ref{eq:twobodyana}), 
we see that $\overline{\kappa}^A$ eliminates the largest three volume
contributions up to relativistic corrections, while 
$\overline{\kappa}^B$ and $\overline{\kappa}^C$ eliminate the first two
contributions, and $\overline{\kappa}^D$ and $\overline{\kappa}^E$ eliminate
only the first.

\section{Numerical Exploration of the Deuteron Binding Energy}
\noindent 
The deuteron is the simplest nucleus, comprised of a neutron and a proton.
Its binding energy is $B=2.224644(34)~{\rm MeV}$ which corresponds to a binding
momentum of $\kappa_0\sim 45.70~{\rm
  MeV}$ (using the isospin averaged nucleon mass of $M_N=938.92~{\rm MeV}$).
As it is a spin-1 system composed of two spin-${1\over 2}$ nucleons, its
wavefunction is an admixture of s-wave and d-wave.  The system is predominantly
s-wave with a small admixture of d-wave induced by the tensor 
($L=S=2$) interaction.
For the purposes of this analysis we will neglect the small d-wave admixture in the deuteron
wavefunction and assume that the deuteron is entirely s-wave.
The low-energy s-wave scattering of a neutron and proton with
$J^\pi=1^+$ is well described by effective range theory, where the ERE of
$p\cot\delta$ converges rapidly with just the first two terms.
The scattering length is known to be $a^{(\siii)}=5.425(1)~{\rm fm}$, the effective
range is  $r^{(\siii)}=1.749(8)~{\rm fm}$,
and   the shape parameter is anomalously small and neglected.
The central values of 
these two parameters  in the s-wave amplitude give rise to a deuteron binding energy
of $B\sim 2.212~{\rm MeV}$ with a  corresponding $\kappa_0$ of 
$\kappa_0\sim 45.58~{\rm MeV}$, which are within $\sim 0.5\%$ of the actual
deuteron binding parameters.

\begin{figure}[!ht]
\begin{tabular}{cc}
\includegraphics[width=0.44\textwidth]{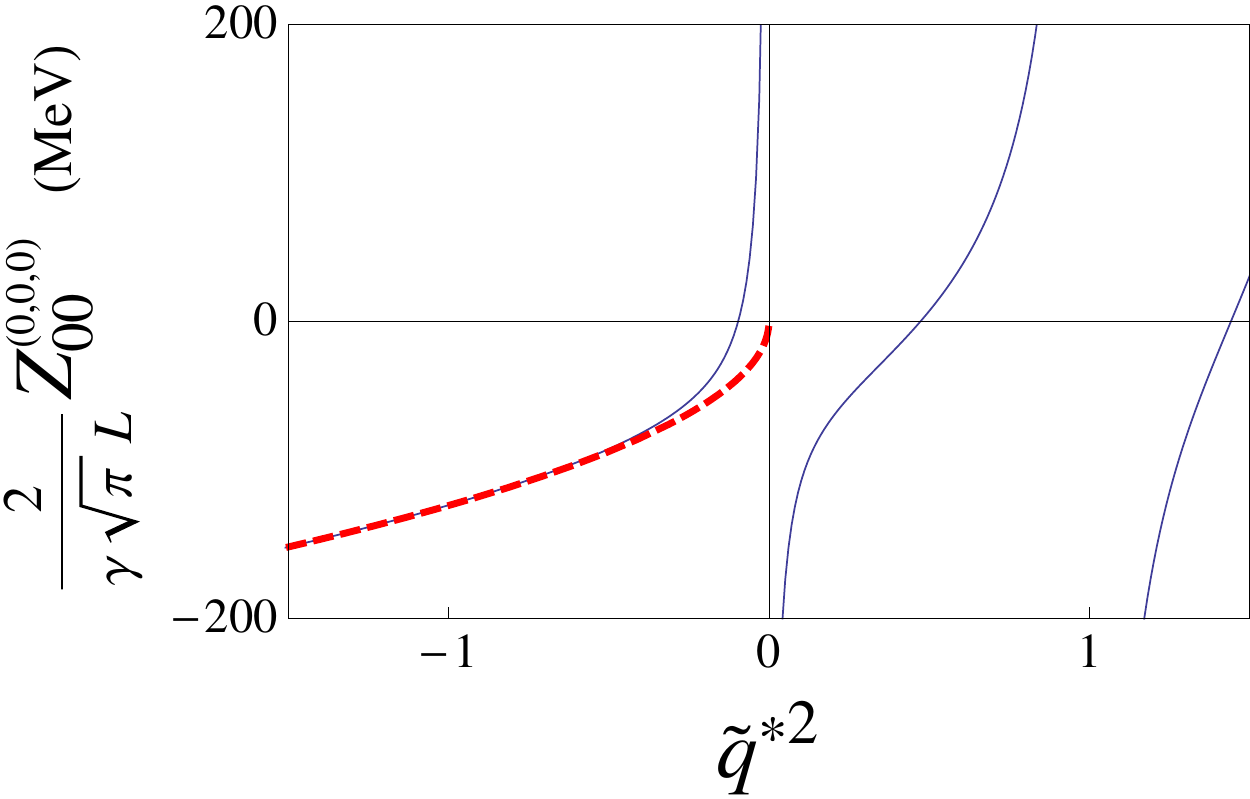}\ \ \ \ \  
\includegraphics[width=0.44\textwidth]{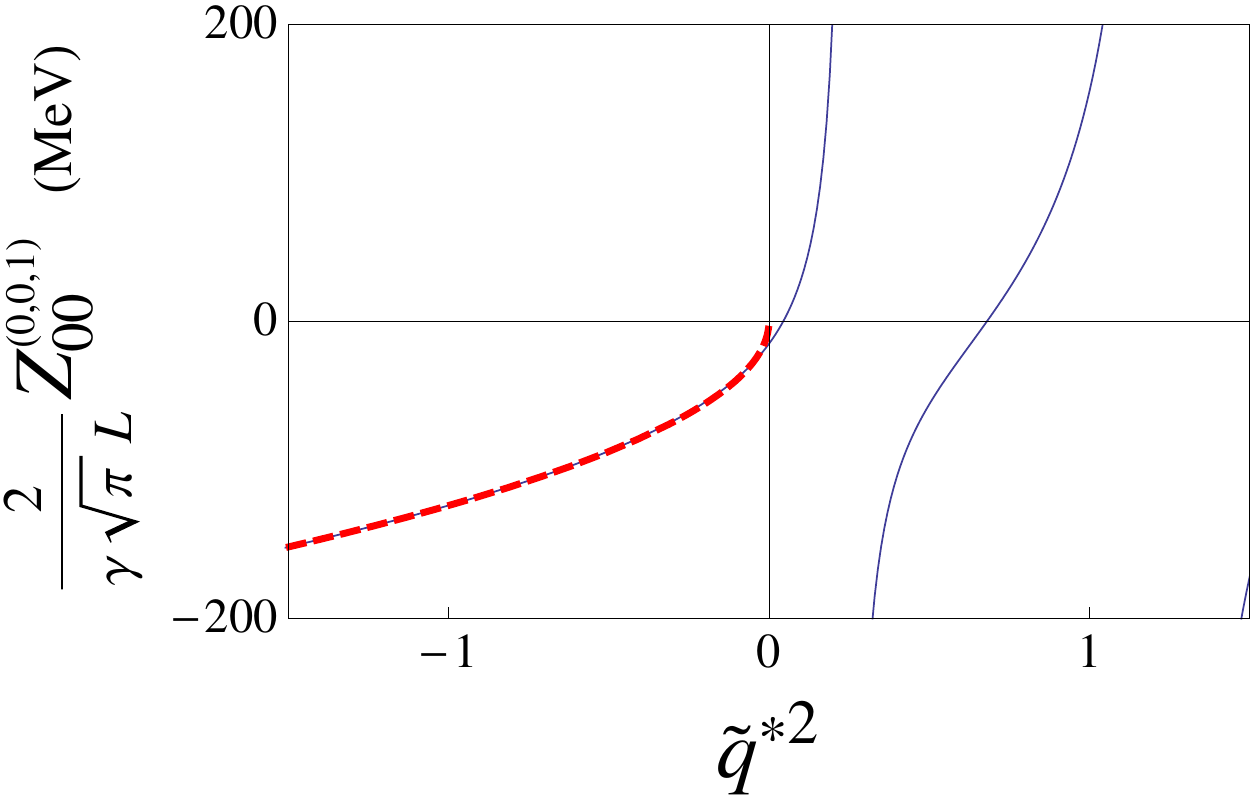}\\ 
\includegraphics[width=0.44\textwidth]{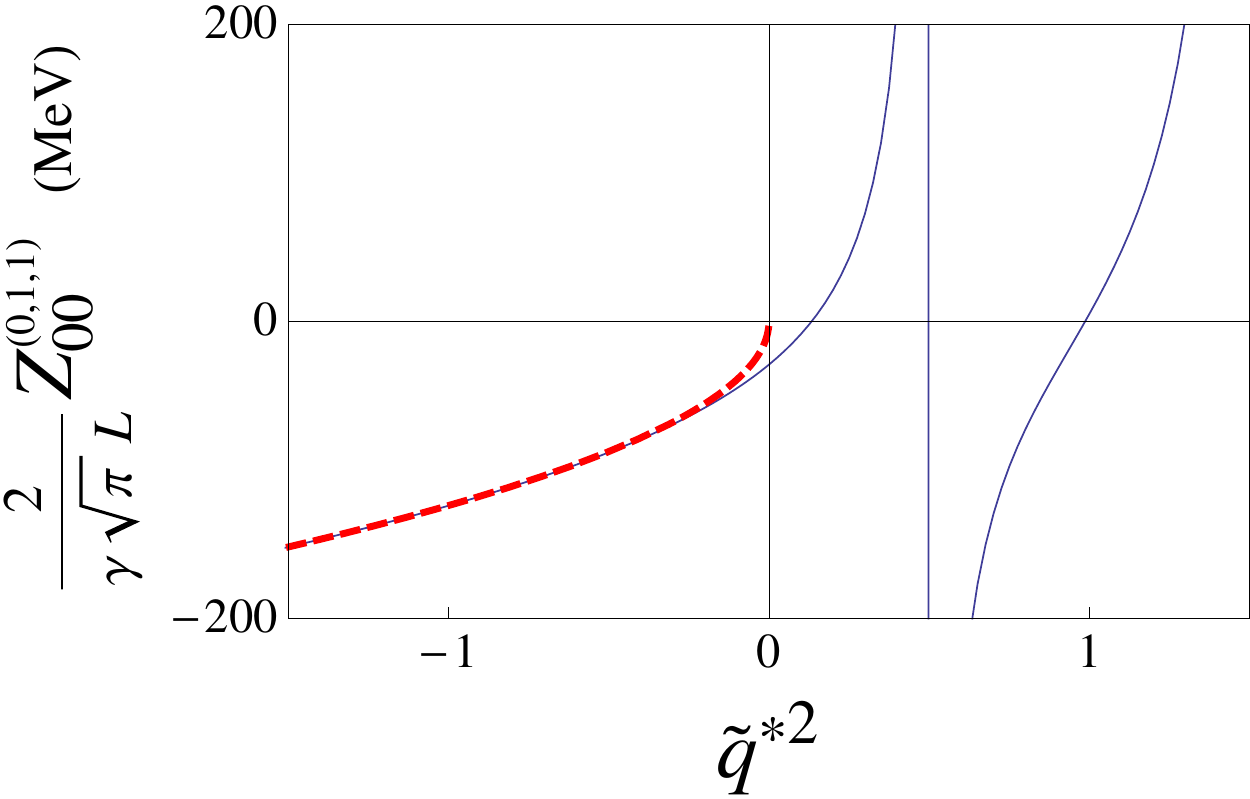}\ \ \ \ \ 
\includegraphics[width=0.44\textwidth]{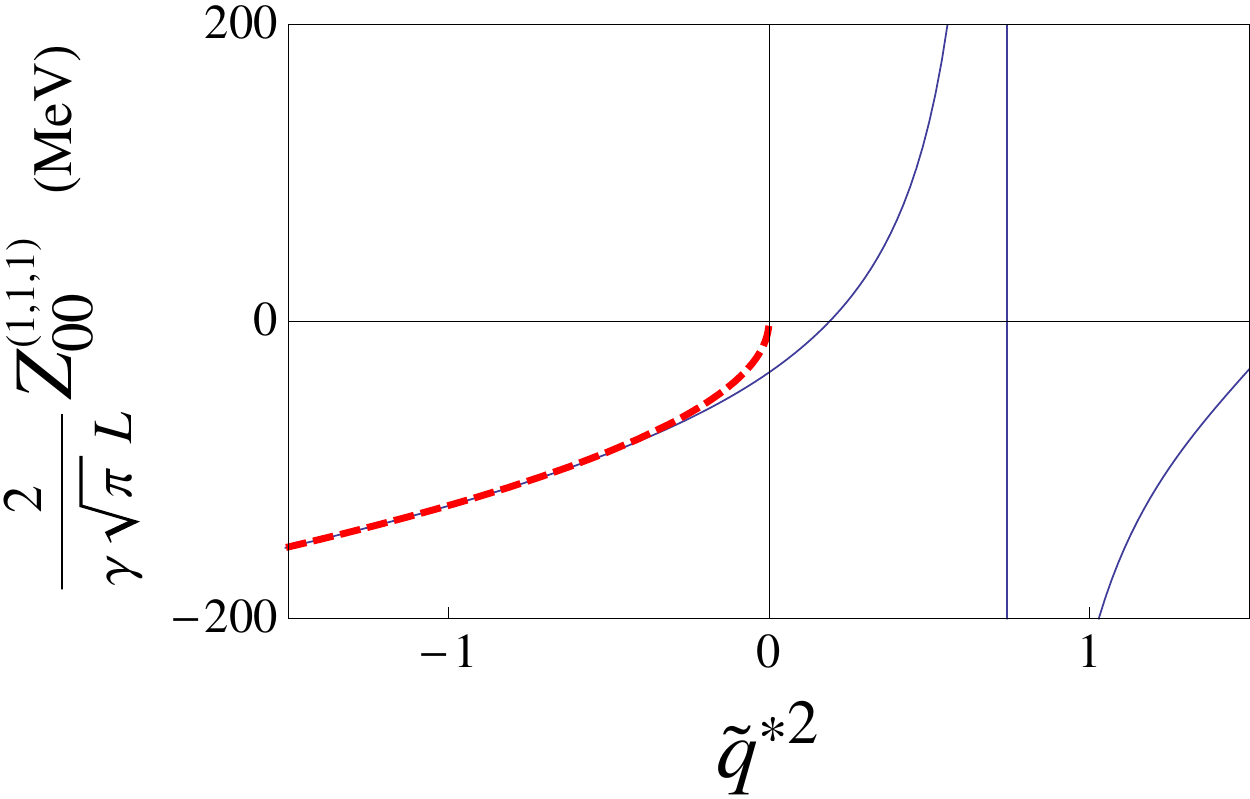} 
\end{tabular}
\caption{\label{fig:Zfuns}
The $Z_{00}^{({\bf d})}(1; \tilde q^{*2},0)$ functions (blue curves), normalized by
${2\over\gamma L \sqrt{\pi}}$, for the deuteron in a cubic volume with 
$L=10~{\rm fm}$.
The dashed (red) curves correspond to $-\sqrt{-\tilde q^{*2}}$,
asymptotic form of the function as $\tilde q^{*2}\rightarrow -\infty$.
The hyperfine splitting of the lowest poles in  the $|{\bf d}|^2=2,3$ functions,
determined by $\gamma-1$, is evident.
}
\end{figure}
\begin{figure}[!ht]
\begin{tabular}{cc}
\includegraphics[width=0.8\textwidth]{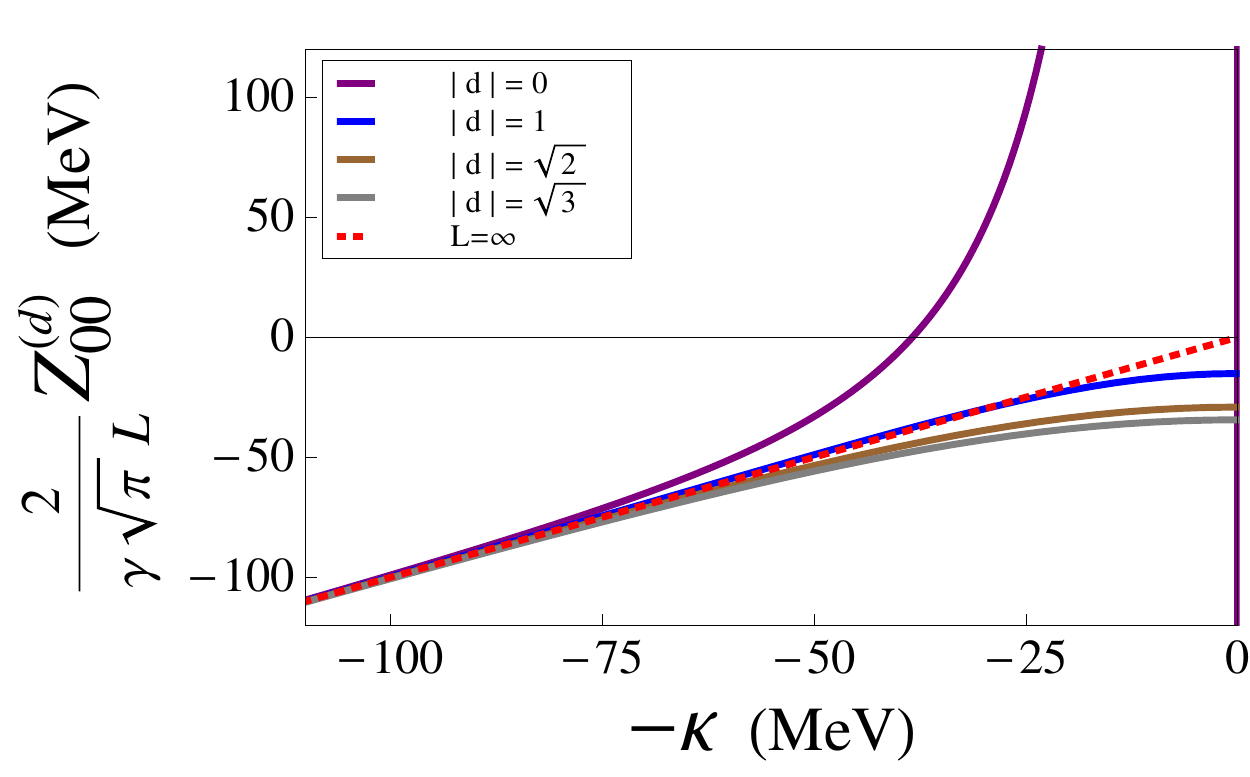}\ \ \ \ \ \ 
\end{tabular}
\caption{\label{fig:Ztails}
The ${2\over\gamma L \sqrt{\pi}} Z_{00}^{({\bf d})}(1; \tilde q^{*2},0)$ 
functions for the deuteron with $|{\bf d}|^2=0,1,2,3$ in a cubic volume with
$L=10~{\rm fm}$
evaluated at  $ \tilde q^{*2} = - \kappa^2 {L^2\over 4\pi^2}$.
The dotted (red) curve corresponds to the asymptotic form of $-\kappa$ for $L\rightarrow\infty$.
}
\end{figure}
\begin{figure}
\begin{tabular}{cc}
\includegraphics[width=0.8\textwidth]{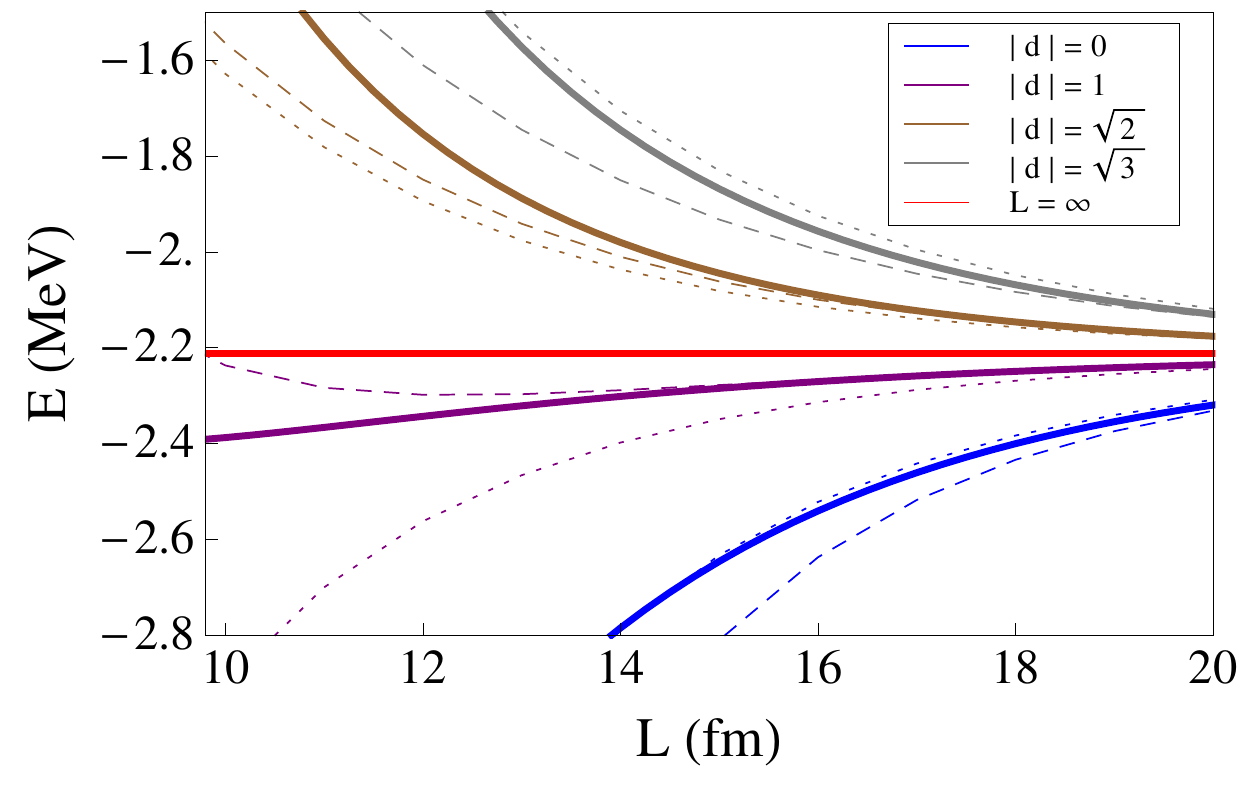}
\end{tabular}
\caption{\label{fig:AnaNum}
The ground state energy in the deuteron channel.  The blue, purple, brown and
gray solid curves are the
exact energy for each system with total momentum defined by $|{\bf d}|$
determined from eq.~(\protect\ref{eq:pcotkappa}).
The solid red lines is the infinite volume ground state energy.
The dotted curves result from using the analytic forms for the volume
modifications at ${\cal O}\left( e^{-\kappa_0 L}/L\right)$, 
given in eq.~(\protect\ref{eq:twobodyana}),
while the 
dashed curves are from the analytic forms up to and including  ${\cal O}\left(
  e^{-\sqrt{3} \kappa_0 L}/L\right)$.
}
\end{figure}

The functions ${2\over\gamma L \sqrt{\pi}}Z_{00}^{({\bf d})}(1; \tilde
q^{*2},0)$
for $|{\bf d}|^2=0,1,2,3$ are shown in fig.~\ref{fig:Zfuns}, along with their
asymptotic form  $-\sqrt{-\tilde q^{*2}}$ at large $-\tilde q^{*2}$.
It is clear that, of the four functions shown, the $|{\bf d}|=0$ function
will give rise to the largest volume modifications to the deuteron binding
energy, as reflected in the deviation from its asymptotic form.
This is because this function has a pole at $\tilde q^{*2}=0$, while the 
 $|{\bf d}|^2=1,2,3$ functions do not.
It is also clear from fig.~\ref{fig:Zfuns}, and magnified in fig.~\ref{fig:Ztails},
that the $|{\bf d}|=1$ function
exhibits the smallest deviations from its asymptotic form and the 
volume modifications to the deuteron binding are expected to be the smallest of
the four considered.
An interesting point to note from fig.~\ref{fig:Ztails} is that despite 
$Z_{00}^{(0,0,1)}$ having volume modifications that start at 
${\cal  O}\left( e^{-\kappa L}/L \right)$, there are significant cancellations between
  all of the exponential contributions, leaving the function very close to its
  asymptotic value over most of the range of $\kappa$.
Figure~\ref{fig:AnaNum} shows the ground state energy in the deuteron channel
(negative of the binding energy) as a function of the spatial extent of the
volume through numerical solution of eq.~(\ref{eq:pcotkappa}).  Also shown in
this figure are the contributions from the ${\cal O}\left( e^{-\kappa_0
    L}/L\right)$ volume modifications, and from the  volume modifications up to
and including ${\cal O}\left(  e^{-\sqrt{3} \kappa_0 L}/L\right)$.

\begin{figure}
\begin{tabular}{cc}
\includegraphics[width=0.8\textwidth]{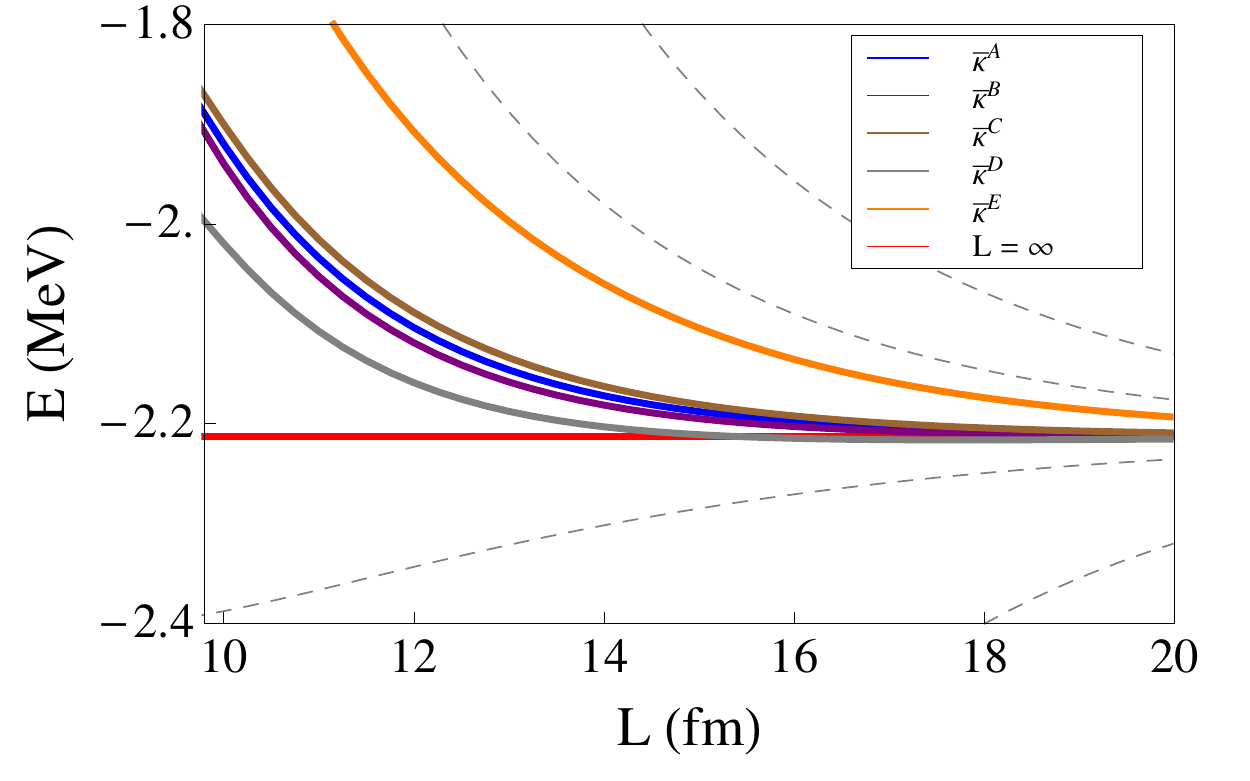}
\end{tabular}
\caption{\label{fig:Money}
The exponentially-improved estimates of the ground state energy in the deuteron
channel from the $\overline{\kappa}^i$ relations given in
eq.~(\protect\ref{eq:twobodyana}).
The gray dashed lines denote the ground state energies in the systems with a
given ${\bf d}$-vector that are shown as the solid curves in fig.~\protect\ref{fig:AnaNum}.
}
\end{figure}

Forming the linear combinations of the $\kappa^{({\bf d})}$
given in eq.~(\ref{eq:twobodyana}), the
$\overline{\kappa}^i$,  from the exact numerical 
solutions to eq.~(\ref{eq:pcotkappa}),
gives rise to the improved estimates of the 
deuteron binding energy 
that are shown in fig.~\ref{fig:Money}.
Surprisingly, there is little difference between the volume
modifications improved to ${\cal O}\left( e^{-2\kappa_0 L}/L\right)$ and those
  improved to ${\cal O}\left( e^{-\sqrt{3}\kappa_0 L}/L\right)$ 
for volumes with $10~{\rm fm} \lsim L \lsim 20~{\rm fm}$.
For $L\gsim 12~{\rm fm}$ the $\overline{\kappa}^i$, except for
$\overline{\kappa}^E$, provide estimates of the deuteron binding energy that
are significantly closer to its actual binding energy than the ground state
of any given ${\bf d}$ spectrum.
The $\overline{\kappa}^D$ combination is closer to the infinite volume binding
energy than one would expect.  While it is improved to ${\cal O}\left(
  e^{-\sqrt{2}\kappa_0 L}/L\right)$ it appears to be better than any of the
others that are improved to higher orders. However, this is true only at these
``intermediate'' volumes, while in the very large volumes the predicted hierarchy
is, in fact,  found. In this combination there is a subtle cancellation between
different volume dependences in the range of volumes that are shown in
fig.~\ref{fig:Money}.

\section{Volume-Improved Fitting}
\noindent 
While it is important to form the exponentially volume-improved combinations
of binding momenta, it  may not be the method that is actually implemented in
the analysis of the results of LQCD calculations.  
The existence of the relations shows that 
the volume modifications in  a prediction of the deuteron binding energy 
from one ensemble of gauge-field configurations can be
exponentially reduced (in the NR-limit) with minimal additional computational
resources.
However, this reduction can also be accomplished simply by fitting the
appropriate volume dependences to the results of the LQCD calculations for a
range of ${\bf d}$.
From eq.~(\ref{eq:Zkappa}), the binding momentum for any given  ${\bf d}$ is
\begin{eqnarray}
\kappa^{({\bf d})} & = & \kappa_0\ +\ {Z_\psi^2\over L}\ F^{({\bf d})}(\kappa_0
L)\ +\ {\cal O}\left( e^{-2\kappa_0 L}/L\right)
\ \ \ ,
\end{eqnarray}
with the coefficients and kinematic factors in $F^{({\bf d})}(\kappa_0 L)$
determined by the lattice calculation. 
Therefore, up to ${\cal O}\left( e^{-2\kappa_0 L}/L\right)$,
the two free-parameters that remain
to be determined are $\kappa_0$ and $Z_\psi$ which can be accomplished with a
$\chi^2$-minimization.  In the case of having 
$\kappa^{({\bf d})}$ for only two  ${\bf d}$'s,  $\kappa_0$ and $Z_\psi$ can be
solved for  within the uncertainties of the LQCD calculations~\footnote{
The volume extrapolation performed in Ref.\cite{Beane:2010hg}
  to determine the H-dibaryon binding energy,  used the ground state energy obtained
  in two different lattice volumes and iteratively solved for  $\kappa_0$ and $Z_\psi$
in $\kappa^{(0,0,0)}$.}.

\section{Conclusions}
\noindent 
As recently stressed by Bour {\it et al}, 
the binding energy of a bound state depends upon its total momentum when
subject to periodic boundary conditions in the spatial directions.  
Through the momentum modes excluded by the boundary conditions,
these volume modifications of the binding energy depend upon the
ratio of the spatial extent of the volume to the Lorentz-contracted size of the
bound state, and also upon the masses of the constituents.
We have extended the work of Bour {\it et al} from nonrelativistic quantum
mechanics to quantum field theory and 
have pointed out that these features can be utilized in Lattice QCD calculations of
hadronic bound states to (approximately exponentially) reduce the volume
modifications of predicted binding energies.
The standard Lattice QCD methodology that is used to determine the binding energy of
a bound state is to measure the ground state energy of a system in a number of
lattice volumes and then extrapolate to $L=\infty$ with a function of the form  $\sim
e^{-\kappa_0 L}/L + {\cal O}\left( e^{-\sqrt{2}\kappa_0 L}/L\right)$.
Using combinations of boosted ground state energies,
the volume dependence of the binding energy can be exponentially reduced in the nonrelativistic
limit.  For instance, 
the ground state energies of the 
lowest four boosted states in the lattice volume can be combined to reduce the volume
modifications to the predicted binding energy to  
$\sim e^{-2 \kappa_0 L}/L + {\cal O}\left( \eta^2 e^{-\kappa_0 L}/L\right)$
where $\eta\ll 1$.

In the specific case of the deuteron (neglecting its d-wave
component), we have numerically explored what might be
expected from future Lattice QCD calculations, and in particular, examined the
volume dependence of combinations of boosted ground state energies.  
We find that the deuteron binding energy can be extracted to high precision in
reasonably modest volumes, reducing the volume modifications by more than an
order of magnitude over those of the state at rest for $L\gsim 10~{\rm fm}$.
It is also found that the volume modifications to the binding energy of the
system with one unit of lattice momentum are significantly smaller than those
of other low-momentum states, including the state at rest.
It is clear that future Lattice QCD calculations that focus on extracting the properties and interactions
of nuclei, including exotic systems such as the H-dibaryon, can greatly
enhance the precision of their predicted binding energies by including
systems with nonzero total momentum into their production.

\vskip0.2in

We would like to thank William Detmold and Dean Lee for useful discussions.
ZD and MJS were supported in part by the DOE grant DE-FG03-97ER4014.

\end{document}